\documentclass{fmj2010}
\usepackage{graphicx}
\usepackage[dvips]{epsfig}
\usepackage{url}
\def\fgrst{\textit{Fermi Gamma-Ray Space Telescope}}
\def\fermilat{\textit{Fermi}/LAT}
\def\fermi{\textit{Fermi}}

\begin{document}
   \title{VLBI Monitoring of the bright $\gamma$-ray blazar PKS\,0537$-$441 
}

   \author{F. Hungwe\inst{1,}\inst{4}
          R. Ojha\inst{2}, M. Kadler\inst{3,}\inst{6,}\inst{7}, R. Booth \inst{4}, J. Blanchard\inst{5}, J. Lovell\inst{5}, C. M\"uller\inst{3}, M. B\"ock\inst{3}   \and the TANAMI team
          }

   \institute{Department of Physics \& Electronics, Rhodes University, PO Box 94, Grahamstown 6140, South Africa
         \and
            NVI/United States Naval Observatory, 3450 Massachusetts Ave, NW, Washington, DC 20392-5420
         \and 
              Dr. Remeis-Sternwarte \& ECAP, Universit\"at Erlangen-N\"urnberg, Sternwartstr.~7, 96049 Bamberg, Germany
       \and 
              Hartebeesthoek Radio Astronomy Observatory, PO Box 443, Krugersdorp 1740, South Africa
     \and
            School of Mathematics \& Physics, Private Bag 37, University of
Tasmania, Hobart TAS 7001, Australia
\and 
         CRESST/NASA Goddard Space Flight Center, Greenbelt, MD 20771, USA
\and
    Universities Space Research Association, 10211 Wincopin Circle, Suite 500 Columbia, MD 21044, USA
             }
\authorrunning{F. Hungwe et. al}

   \abstract{One of the defining characteristics of BL Lacertae objects is their strong variability across the electromagnetic spectrum. PKS 0537$-$441 is one such object and is one of the most luminous blazars from radio to $\gamma$-ray wavelengths. It was detected as a strong and highly variable source by EGRET and has been reported several times to be in an active state by \fermi . It is one of the brightest $\gamma$-ray blazars detected in the southern sky so far. 
The TANAMI (Tracking Active Galactic Nuclei with Austral Milliarcsecond Interferometry) program is monitoring PKS 0537$-$441 at VLBI resolutions. We present 8.4\,GHz and 22\,GHz images of the milliarcsecond scale structure. We also present our ongoing analysis of the spectral and temporal changes in this object.}

   \maketitle
%

\begin{figure}
   \centering
   \includegraphics[width=8.7cm]{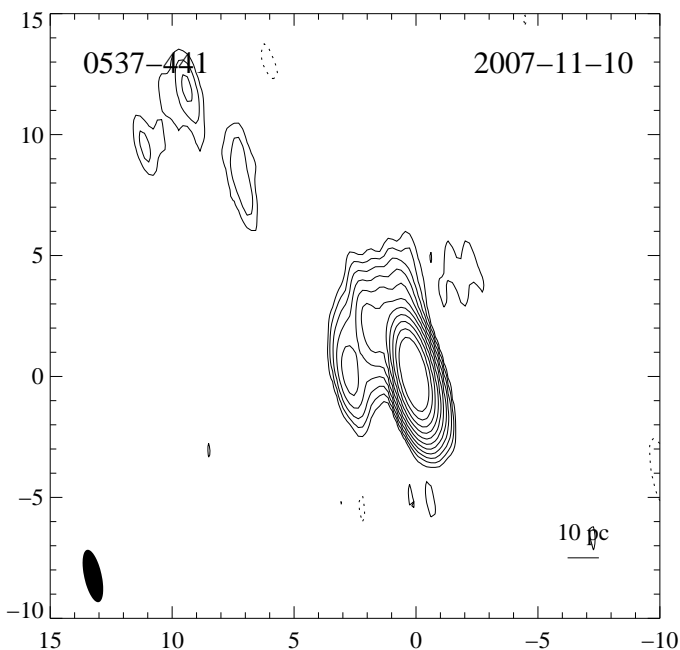}\\
      \caption{$8.4$\,GHz radio images for PKS 0537$-$441 for observations done in Nov 2007. x- and y-axis are labeled in milliarcseconds and the hatched ellipse on the bottom left represents the beam of the observing array.}
   \end{figure}

\section{Introduction}

PKS\,0537$-$441 is a strongly variable BL Lacertae object  and is one of the brightest $\gamma$-ray blazars detected in the southern sky to date. The source is known to be a strong intra-day variable and is sometimes classified as a highly polarised quasar (\cite{Treves:93}). It has been a candidate for gravitational microlensing (\cite{Romero:95}) but \cite{Heidt:03} disprove this and suggest it might be part of binary quasar. 

Using the Australian Long Baseline Array (LBA) and affiliated telescopes, the Tracking Active Galactic Nuclei with Austral Milliarcsecond 
Interferometry (TANAMI) program (\cite{Ojha2010}) has been monitoring southern sky blazars such as PKS\,0537$-$441 at VLBI resolutions. TANAMI began observing known and possible $\gamma$-ray loud extragalactic sources south of -30 degrees about a year before the launch of \fermi.  TANAMI observations at $8.4$\,GHz and $22$\,GHz are made about every 2\, months at milliarcsecond resolution. The TANAMI sample is modified to include 
sources within this declination range that are detected by the Large Area Telescope (LAT) on board \fgrst. 

Studying Active Galactic Nuclei (AGN) at different wavelengths is crucial in order to understand AGN-jets and differentiate between different models explaining the jets. TANAMI observations of PKS\,0537$-$441 complement ongoing observations at other wavelengths. This source is on the \fermilat ~monitored source list and is also in the LAT 1-year Point Source Catalog (\cite{Abdo2010c}).

PKS\,0537$-$441 has been reported by \fermilat ~to be active about four times in the past two years. In October 2008 and July 2009, the \fermilat  ~observed an increase in $\gamma$-ray activity in the source (\cite{ATEL:1759}) \& (\cite{ATEL:2124}). In February of 2010, AGILE detected `enhanced $\gamma$-ray emission above 100\,MeV from a source positionally consistent with the blazar PKS\,0537$-$441' (\cite{ATEL:2454}). Most recently (April 2010) the \fermilat  ~observed increased $\gamma$-ray activity in this source. 

\begin{figure*}
  \begin{center} 
    \begin{tabular}{c}
\epsfig{file=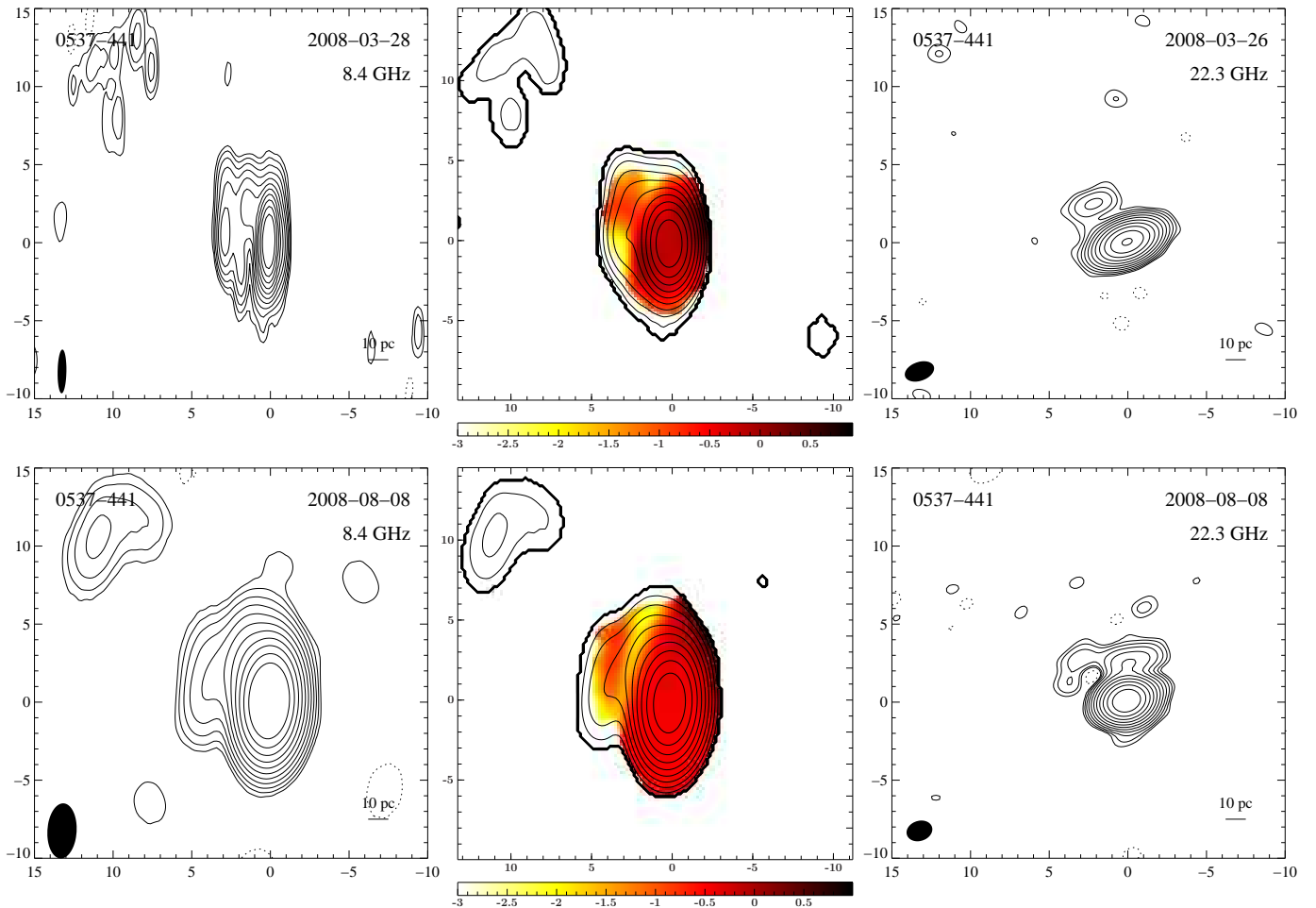,width=18cm,angle=0}
 \end{tabular}
\caption{Images from 2 epochs of dual frequency observations of PKS\,0537$-$441. x- and y-axis are labeled in milliarcseconds and the hatched ellipse on the bottom left represents the beam of the observing array. In each row, the left and right images are observations at $8.4$\,GHz and 22\,GHz respectively. The image at the centre of each row is the spectral index map made from the images on either side. The colour coding shows the spectral index as defined in the text.}
  \end{center}
\end{figure*}


\section{Observations and Data Reduction}

VLBI observations for the epochs presented were made using the Australian Long Baseline Array (LBA\footnote{The Long Baseline Array is part of the Australia Telescope which is funded by the Commonwealth of Australia for operation as a National Facility managed by CSIRO.}) together with the Hartebeesthoek 26m dish in South Africa. The LBA consists of five telescopes, namely, Parkes ($64$\,m), Narrabri ($5$x$22$\,m), Ceduna ($30$\,m), Hobart ($26$\,m) and Mopra ($22$\,m). In addition, TANAMI sometimes has access to two other telescopes within Australia, the $70$\,m and $34$\,m telescopes belonging to NASA's Deep Space Network at Tidbinbilla. These observations were made at $8.4$\,GHz and $22$\,GHz. Each observing run is $24$\,hours long with every source being observed for about an hour. The observations 
are made approximately every 2\,months. The data were correlated at the Swinburne University Correlater. The radio data were calibrated using AIPS and the radio images were made using the CLEAN algorithm in \textsc{difmap} interactive mode (\cite{Shepherd:97}). For details of the observations, calibration and imaging see \cite{Ojha2010}. Here we present VLBI images from three epochs of observations two of which have data both at X and K-band. 

The data used for the radio light curve was taken using the University of Tasmania's $30$\,m telescope. This dish, together with the $26$\,m dish at Hobart, have since August 2007, been monitoring TANAMI sources (PI: Jim Lovell). Hobart observes at $2.3$\,GHz and $6.4$\,GHz while Ceduna observes at $6.7$\,GHz. Weekly, $48$\,hour observations are done with about one hour spent observing a single source for every epoch.

The $\gamma$-ray observations were made using the LAT. The \fermilat ~has a number of observing modes but spends most of its time in survey modes where the whole sky is scanned once every two orbits. For each event, the \fermilat ~measures 3 quantities, the arrival direction, the energy and the arrival time. All \fermilat\ data shown here were downloaded from the public website\footnote{\url{http://fermi.gsfc.nasa.gov/cgi-bin/ssc/LAT/LATDataQuery.cgi}}. The data were reduced using \fermi ~science tools\footnote{\fermi ~science tools can be downloaded from \url{http://fermi.gsfc.nasa.gov/ssc/data/analysis/software/}} using monthly time bins and an energy range of $100$\,MeV to $300$\,GeV. GTSELECT was used to specify the time (in mission elapsed time) and energy range and the region of interest. To make good time intervals, removing data when the \fermi ~spacecraft was over the South Atlantic Anomaly, GTMKTIME was used. Livetime cubes were calculated using  GTLTCUBE while GTEXPMAP was used to generate exposure maps for the region of interest. GTLIKE was then used to perform the likelihood analysis of the \fermilat ~data. The output of GTLIKE gives the integrated flux and the test statistic among other parameters.

\section{TANAMI images of PKS\,0537$-$441}

Figure 1 shows an $8.4$\,GHz image from November 2007.
Figure 2 shows $8.4$\,GHz and $22$\,GHz radio images from observations at two additional epochs  (March and August 2008). In each row the X- and K-band images are shown at the left and right, respectively. The corresponding spectral index map made using these images is shown in between the images. Details of all these images are summarized in Table 1. 

\begin{table*}
\caption{Image parameters and observation characteristics}
\label{table:1} 
\centering 
\begin{tabular}{c c c c c c c c}
\hline\hline 
Frequency & Epoch & RMS & $S_{\mathrm{peak}}$ & $S_{\mathrm{total}}$ & $\theta_{\mathrm{maj}}$ &  $\theta_{\mathrm{min}}$ & P.A. \\
$[\mathrm{GHz}]$    & yyyy-mm-dd& [$\mathrm{mJy\,beam^{-1}}$] & [$\mathrm{Jy\,beam^{-1}}$] & [$\mathrm{Jy}$] & [$\mathrm{mas}$] & [$\mathrm{mas}$] & [$^\circ$]\\
\hline 
8.4 & 2007-11-10 &0.37    $\pm 0.40$ &  4.85       &    5.10   &  2.2 & 0.7& 12 \\ 
8.4 & 2008-03-28 &0.58 $\pm  0.11 $ &    3.38   &       4.24  &    2.8 & 0.6 & -1     \\ 
8.4 & 2008-08-08 & 0.39    $\pm0.35$    &   4.67       &    4.91 &    3.5 & 1.9 &-3    \\ 
22.3 & 2008-03-26 &0.47 $\pm 0.99$ &    3.12    &       3.43    & 1.9 &  1.2 & -70     \\
22.3 & 2008-08-08 &0.47 $\pm 0.36$  &   3.00    &   3.21    & 1.7 & 1.3& -72 \\
\hline 
\end{tabular}\\
\end{table*}

The $8.4$\,GHz images show a radio core and extended emission to the east. The images show more extended emission in March 2008 
and even more in August 2008. This is the time leading to the $\gamma$-ray flare observed weeks leading to October 2008. The core is unresolved at $22$\,GHz for both epochs. Observations at the two frequencies allow calculations of the spectral indices which, when used together with indices from other energy bands, e.g. $\gamma$-ray, give the broadband Spectral Energy Distribution (SEDs) needed to understand Active Galactic Nuclei.

The mean flux density $\bar{S}$ at the $8.4$\,GHz epochs is $\bar{S}_\textnormal{peak} \approx 4.3$\,Jy beam$^{-1}$ while $\bar{S}_\textnormal{total} \approx 4.75$\,Jy. The peak flux density $S_\textnormal{peak}$ at $8.4$\,GHz shows significant variability in the flux density with a standard deviation of $0.80$\,Jy beam$^{-1}$,  while the total flux density $S_\textnormal{total}$ has a standard deviation of $0.45$\,Jy. At 22\,GHz, there is a slight decrease in the flux density from the March 2008 epoch to the August 2008 epoch.

Figure 2 shows the spectral index maps derived from the two observing frequencies, $8.4$\,GHz and $22$\,GHz. The spectral index  is defined as \mbox{$F_\nu \sim \nu^\alpha$}.  The spectral index maps  of PKS 0537$-$441 were calculated according to this definition where $S_{\mathrm{8.4GHz}}\geq 5\sigma_{\mathrm{8.4GHz}}$ and $S_{\mathrm{22GHz}}\geq 5\sigma_{\mathrm{22GHz}}$.  The color coding shows the changing of the spectral index from optically thin to optically thick emission regions. The overlaid contours show the flux density distribution at 8.4\,GHz folded with the beam of the corresponding 22\,GHz image. The core of the source has an almost perfectly flat spectrum, while the jet has an optically thin spectrum with a hint of steepening with time. As we have not modeled any optically thin components, we have not been able to calculate a core shift for PKS\,0537$-$441
yet.

\section{Radio and $\gamma$-ray Light Curves of PKS\,0537-441}
\begin{figure}[!h]
   \centering
   \includegraphics[width=6.9cm,angle=-90]{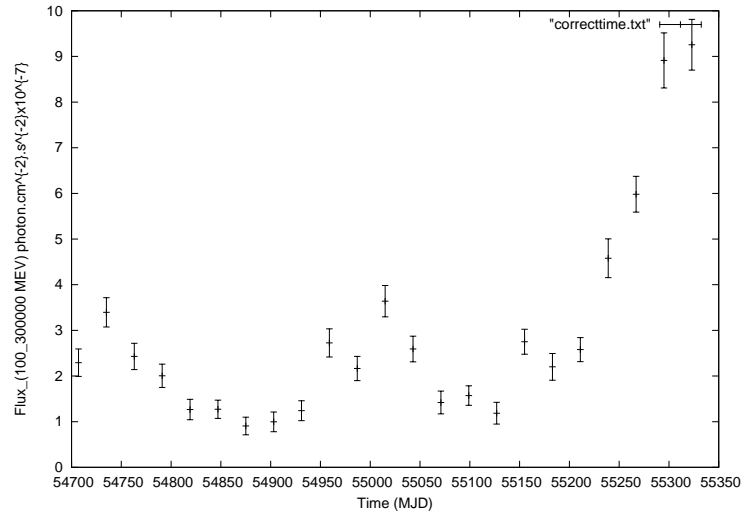}
      \caption{Gamma-ray light curve for PKS 0537$-$441 spanning $\sim$21 months from August 2008 to May 2010.}
   \end{figure}

Figures 3 and 4 show the $\gamma$-ray light curve for data from August 2008 to May 2010 and the radio light curve for the period 14 April to 28 May 2010, respectively. The points on the $\gamma$-ray light curve represent monthly bin sizes. Photons with energies from $100$\,MeV to $300$\,GeV were considered.  

PKS\,0537$-$441 has been reported by the Astronomer's Telegram to be in an active state in October 2008, July 2009, February 2010 and April 2010. The $\gamma$-ray light curve shows the source to be active in September 2008, July 2009 and April 2010. When reducing the data, we used monthly bin sizes and this is a possible source of the disparity. Also, most telegrams are sent at the onset of increased activity and not necessarily the peak of such activity.

\begin{figure}[!h]
   \centering
   \includegraphics[width=6.9cm,angle=-90]{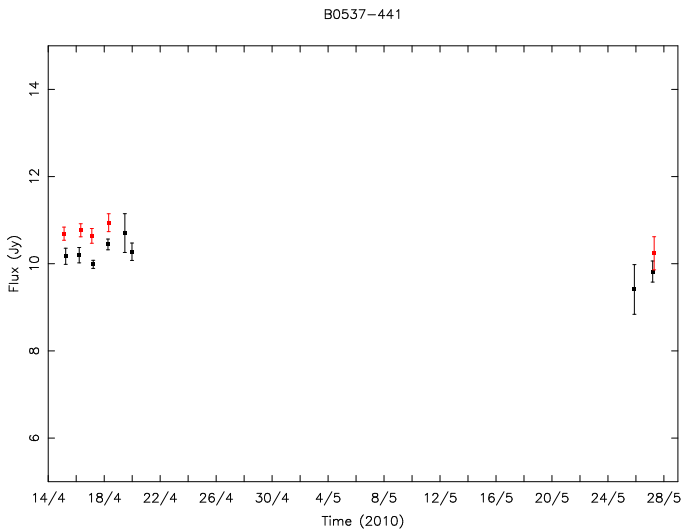}
      \caption{Radio light curve for PKS 0537$-$441. Note that these data cover a much shorter time compared to the gamma-ray light curve.}
   \end{figure}

The data for the radio light curve was taken at $6.7$\,GHz using the University of Tasmania's 30m telescope at Ceduna outside of Adelaide, Australia. This (when not observing as part of the LBA and other VLBI observations) monitors ~15 sources continuously, with an additional ~50 observed once every two weeks.
The continuous monitoring is split into two groups, a southern group and a northern group, which are alternated in two week blocks (though there is some overlap).

Data is taken in four scan blocks, two scans in RA, and two in DEC and fit with a gaussian on a 2nd order polynomial baseline. The two scans in each direction are merged, and cross checked. If the amplitudes do not match, the entire four scan block is rejected. A pointing offset and correction is then calculated and applied in both RA and DEC. The resulting RA and DEC, pointing corrected, scans are also then cross checked and merged. Again the entire block is rejected if the RA and DEC amplitudes don't match.

The resulting amplitude is then corrected for gain-elevation effects and calibrated. The northern group uses 3C227 as a calibrator and the southern B1934-638. Data finally is binned into one day bins (unless intra-day scintillation observations are being made).

The radio light curve covers a much shorter time period than the $\gamma$-ray curve and does not show evidence for variability. We will continue to monitor this source in the radio to look for possible connections between the radio and $\gamma$-ray lightcurves. 


\section{Conclusion}

Multi-epoch TANAMI observations of the highly luminous blazar PKS\,0537$-$441 at two radio frequencies and single dish monitoring data from the Ceduna radio telescope are presented here in addition to a $\gamma$-ray light curve from \fermilat. We have used simultaneous observations at $8.4$ and $22$\,GHz to produce a spectral index map that appears to be the first for this source at these resolutions. These data are being combined with simultaneous data from \fermi\ to construct SEDs and understand the behaviour of this interesting object. Modeling of multiple epochs of data in order to understand the kinematic behaviour is also in progress.

\begin{acknowledgements}
FH acknowledges support from the South African SKA Project and Hartebeesthoek Radio Astronomy Observatory. 
This research has been partially funded by the Fermi Guest Investigator 
Program. 
This research has been partially funded by the Bundesministerium f\"ur
Wirtschaft und Technologie under Deutsches Zentrum f\"ur Luft- und
Raumfahrt grant number 50OR0808.  
\end{acknowledgements}

\end{document}